\documentclass{article}
\usepackage{microtype}
\usepackage[english]{babel}
\usepackage{spconf}
\usepackage{amsmath,amsfonts,mathtools}
\usepackage{siunitx}
\usepackage{graphicx}
\usepackage[dvipsnames]{xcolor}
\usepackage{blindtext}
\usepackage{cite}
\usepackage[moderate,title=normal,sections=normal,margins=normal,bibliography=normal,mathspacing=tight,mathdisplays=tight]{savetrees}
\usepackage[nopdftrailerid=true]{pdfprivacy}
\usepackage{pgfplots}
\usepackage{tikz}
\pgfplotsset{compat=newest}
\pgfplotsset{plot coordinates/math parser=false} 
\usepgfplotslibrary{patchplots}
\usepgfplotslibrary{colormaps}%
\pgfplotsset{%
	colormap/inferno/.style={
		colormap name=inferno,
	},
}
\pgfplotsset{
colormap={inferno2}{
	rgb=(0.001462, 0.000466, 0.013866)
	rgb=(0.037668, 0.025921, 0.132232)
	rgb=(0.116656, 0.047574, 0.272321)
	rgb=(0.217949, 0.036615, 0.383522)
	rgb=(0.316282, 0.053490, 0.425116)
	rgb=(0.410113, 0.087896, 0.433098)
	rgb=(0.503493, 0.121575, 0.423356)
	rgb=(0.596940, 0.154848, 0.398125)
	rgb=(0.688653, 0.192239, 0.357603)
	rgb=(0.775059, 0.239667, 0.303526)
	rgb=(0.851384, 0.302260, 0.239636)
	rgb=(0.912966, 0.381636, 0.169755)
	rgb=(0.956852, 0.475356, 0.094695)
	rgb=(0.981895, 0.579392, 0.026250)
	rgb=(0.987464, 0.690366, 0.079990)
	rgb=(0.973088, 0.805409, 0.216877)
	rgb=(0.947594, 0.917399, 0.410665)
	rgb=(0.988362, 0.998364, 0.644924)
},
}
\DeclarePairedDelimiter\curlyBrace\{\}
\newcommand{\myExpectation}[1]{\operatorname{\mathcal{E}}\curlyBrace*{#1}}
\newcommand{\myPrincipalEigenvec}[1]{\operatorname{\mathcal{P}}\curlyBrace*{#1}}
\newcommand{\numLoc}{M_{\rm H}}
\newcommand{\numEmics}{M_{\rm E}}
\newcommand{\refSpeechDP}{X_{1}^{\rm DP}\left(l\right)}
\newcommand{\vectorY}{\mathbf{y}\left(l\right)}
\newcommand{\vectorYherm}{\mathbf{y}^{H}\left(l\right)}

\newcommand{\vectorX}{\mathbf{x}\left(l\right)}
\newcommand{\vectorXdp}{\mathbf{x}^{\rm DP}\left(l\right)}
\newcommand{\vectorXrev}{\mathbf{x}^{\rm R}\left(l\right)}
\newcommand{\RTFvec}{\mathbf{g}}
\newcommand{\RTFvecherm}{\mathbf{g}^{H}}
\newcommand{\RTFvecHA}{\mathbf{g}_{\rm{H}}\left(\theta\right)}
\newcommand{\RTFvecHANOTHETA}{\mathbf{g}_{\rm{H}}}
\newcommand{\RTFvecE}{\mathbf{g}_{\rm{E}}\left(\theta_{\rm{E}}\right)}
\newcommand{\gProtoHA}{\bar{\mathbf{g}}_{\rm{H}}\left(k,\theta_{i}\right)}
\newcommand{\gProtoHAE}{\bar{\mathbf{g}}_{\rm{H}}\left(k,\theta_{\mathrm{E},j}\right)}

\newcommand{\gProtoE}{\bar{\mathbf{g}}_{\rm{E}}\left(k,\theta_{\mathrm{E},j}\right)}

\newcommand{\gProto}{\bar{\mathbf{g}}^{\rm{match}}\left(k,\theta_{i},\theta_{\mathrm{E},i}\right)}
\newcommand{\gProtoTwoD}{\bar{\mathbf{g}}^{\mathrm{joint}}\left(k,\theta_{i},\theta_{\mathrm{E},j}\right)}
\newcommand{\hatRTFvec}{\hat{\mathbf{g}}\left(k,l\right)}
\newcommand{\hatRTFvecHAkl}{\hat{\mathbf{g}}_{\rm{H}}\left(k,l\right)}

\newcommand{\hatRTFvecHACW}{\hat{\mathbf{g}}_{\rm{H}}^{\left(\rm{CW}\right)}\left(l\right)}
\newcommand{\hatRTFvecCWE}{\hat{\mathbf{g}}^{\left(\rm{CW-E}\right)}\left(l\right)}
\newcommand{\hatRTFvecCWEkl}{\hat{\mathbf{g}}^{\left(\rm{CW-E}\right)}\left(k,l\right)}
\newcommand{\hatRTFvecHACWE}{\hat{\mathbf{g}}_{\rm{H}}^{\left(\rm{CW-E}\right)}\left(l\right)}
\newcommand{\hatRTFvecHACWEkl}{\hat{\mathbf{g}}_{\rm{H}}^{\left(\rm{CW-E}\right)}\left(k,l\right)}
\newcommand{\hatRTFvecECWE}{\hat{\mathbf{g}}_{\rm{E}}^{\left(\rm{CW-E}\right)}\left(l\right)}
\newcommand{\hatRTFvecECWEkl}{\hat{\mathbf{g}}_{\rm{E}}^{\left(\rm{CW-E}\right)}\left(k,l\right)}
\newcommand{\tildeRTFkl}{\tilde{\mathbf{g}}\left(k,l\right)}
\newcommand{\vectorU}{\mathbf{u}\left(l\right)}
\newcommand{\vectorUherm}{\mathbf{u}^{H}\left(l\right)}

\newcommand{\vectorN}{\mathbf{n}\left(l\right)}
\newcommand{\phiY}{\boldsymbol{\Phi}_{\rm y}\left(l\right)}
\newcommand{\phiYkl}{\boldsymbol{\Phi}_{\rm y}\left(k,l\right)}
\newcommand{\phiYHA}{\boldsymbol{\Phi}_{\rm{y,H}}\left(l\right)}
\newcommand{\hatphiY}{\hat{\boldsymbol{\Phi}}_{\rm y}\left(l\right)}
\newcommand{\hatphiYkl}{\hat{\boldsymbol{\Phi}}_{\rm y}\left(k,l\right)}
\newcommand{\hatphiYklprev}{\hat{\boldsymbol{\Phi}}_{\rm y}\left(k,l-1\right)}
\newcommand{\hatphiYHA}{\hat{\boldsymbol{\Phi}}_{\rm{y,H}}\left(l\right)}
\newcommand{\phiU}{\boldsymbol{\Phi}_{\rm u}\left(l\right)}
\newcommand{\phiUkl}{\boldsymbol{\Phi}_{\rm u}\left(k,l\right)}
\newcommand{\phiUHA}{\boldsymbol{\Phi}_{\rm{u,H}}\left(l\right)}
\newcommand{\hatphiU}{\hat{\boldsymbol{\Phi}}_{\rm u}\left(l\right)}
\newcommand{\hatphiUkl}{\hat{\boldsymbol{\Phi}}_{\rm u}\left(k,l\right)}
\newcommand{\hatphiUklprev}{\hat{\boldsymbol{\Phi}}_{\rm u}\left(k,l-1\right)}
\newcommand{\hatphiUHA}{\hat{\boldsymbol{\Phi}}_{\rm{u,H}}\left(l\right)}
\newcommand{\selectionVecE}{\mathbf{e}_{1,\rm{E}}}
\newcommand{\selectionMatrixHA}{\mathbf{E}_{\rm{H}}}
\newcommand{\selectionMatrixE}{\mathbf{E}_{\rm{E}}}

\title{Assisted RTF-Vector-Based Binaural Direction of Arrival Estimation Exploiting a Calibrated External Microphone Array}
\name{Daniel Fejgin and Simon Doclo\thanks{This work was funded by the Deutsche Forschungsgemeinschaft (DFG, German Research Foundation) under Germany's Excellence Strategy - EXC 2177/1 - Project ID 390895286 and Project ID 352015383 - SFB 1330 B2.}}
\address{University of Oldenburg, Department of Medical Physics and Acoustics\\ and Cluster of Excellence Hearing4all, Oldenburg, Germany}
\begin{document}
\ninept
\setlength{\abovedisplayskip}{7pt}
\setlength{\belowdisplayskip}{5pt}
\maketitle
\begin{abstract}
Recently, a relative transfer function (RTF) vector-based method has been proposed to estimate the direction of arrival (DOA) of a target speaker for a binaural hearing aid setup, assuming the availability of external microphones. This method exploits the external microphones to estimate the RTF vector corresponding to the binaural hearing aid and constructs a one-dimensional spatial spectrum by comparing the estimated RTF vector against a database of anechoic prototype RTF vectors for several directions. In this paper, we assume the availability of a calibrated array of external microphones, which is characterized by a second database of anechoic prototype RTF vectors. We propose a method where the external microphones are not only exploited for RTF vector estimation but also assist in estimating the DOA of the target speaker. Based on the estimated RTF vector for all microphones and the prototype RTF databases of the binaural hearing aid and the external microphone array, a two-dimensional spatial spectrum is constructed from which the DOA is estimated. Experimental results for a reverberant environment with diffuse-like noise show that assisted DOA estimation outperforms DOA estimation where the prototype database characterizing the external microphone array is not used.
\end{abstract}
% Abstract: 100 to 150 words
%
\section{Introduction}
\label{sec:intro}
\vskip-1.5ex With the advent of mobile devices that are equipped with microphones, wirelessly linking hearing aids to these devices has become increasingly popular \cite{Mecklenburger2016}. In such an acoustic sensor network, jointly processing all available microphones, i.e. the hearing aid microphones in conjunction with the external microphones, has been shown to be beneficial for, e.g., noise reduction \cite{Bertrand2009,Yee2018,Ali2019,Goessling2019,Goessling2021} as well as for direction of arrival (DOA) estimation \cite{Fejgin2021,Farmani2018,Kavalekalam2019}. 

Large research effort has already been dedicated to DOA estimation \cite{Schmidt19886,DiBiase2001,Gannot2019,Evers2020,Grumiaux2022}, in particular also for binaural hearing aid appli\-cations \cite{May2011,Kayser2014,Fejgin2022,Fejgin2021,Farmani2018,Kavalekalam2019}. In \cite{Fejgin2021} we proposed a DOA estimation method for a binaural hearing aid setup, which exploits an external microphone at an unknown position. In this method all available microphone signals are used to estimate the so-called relative transfer function (RTF) vector between all microphones and a reference microphone on one of the hearing aids. The estimated RTF vector corresponding to only the hearing aid microphones is then compared against a database of anechoic prototype RTF vectors (e.g., obtained through measurement), by constructing a one-dimensional spatial spectrum for different directions. It should be noted that the element in the estimated RTF vector corres\-ponding to the external microphone cannot be used in this comparison, since the position of the external microphone is generally unknown. The DOA of the target speaker relative to the binaural hearing aid setup is then estimated as the direction for which the spatial spectrum is maxi\-mized. Contrary to \cite{Farmani2018}, in \cite{Fejgin2021} the external microphone does not need to be in the vi\-ci\-nity of the target speaker in order to capture a nearly clean speech reference, while contrary to \cite{Kavalekalam2019} no pre-trained representation of clean speech spectral envelopes is required for DOA estimation.
\begin{figure}[b]
	\vskip-5.5mm
	\centering
	\includegraphics[width=0.675\columnwidth,trim={0.5cm 11.5cm 0.5cm 4.5cm},clip]{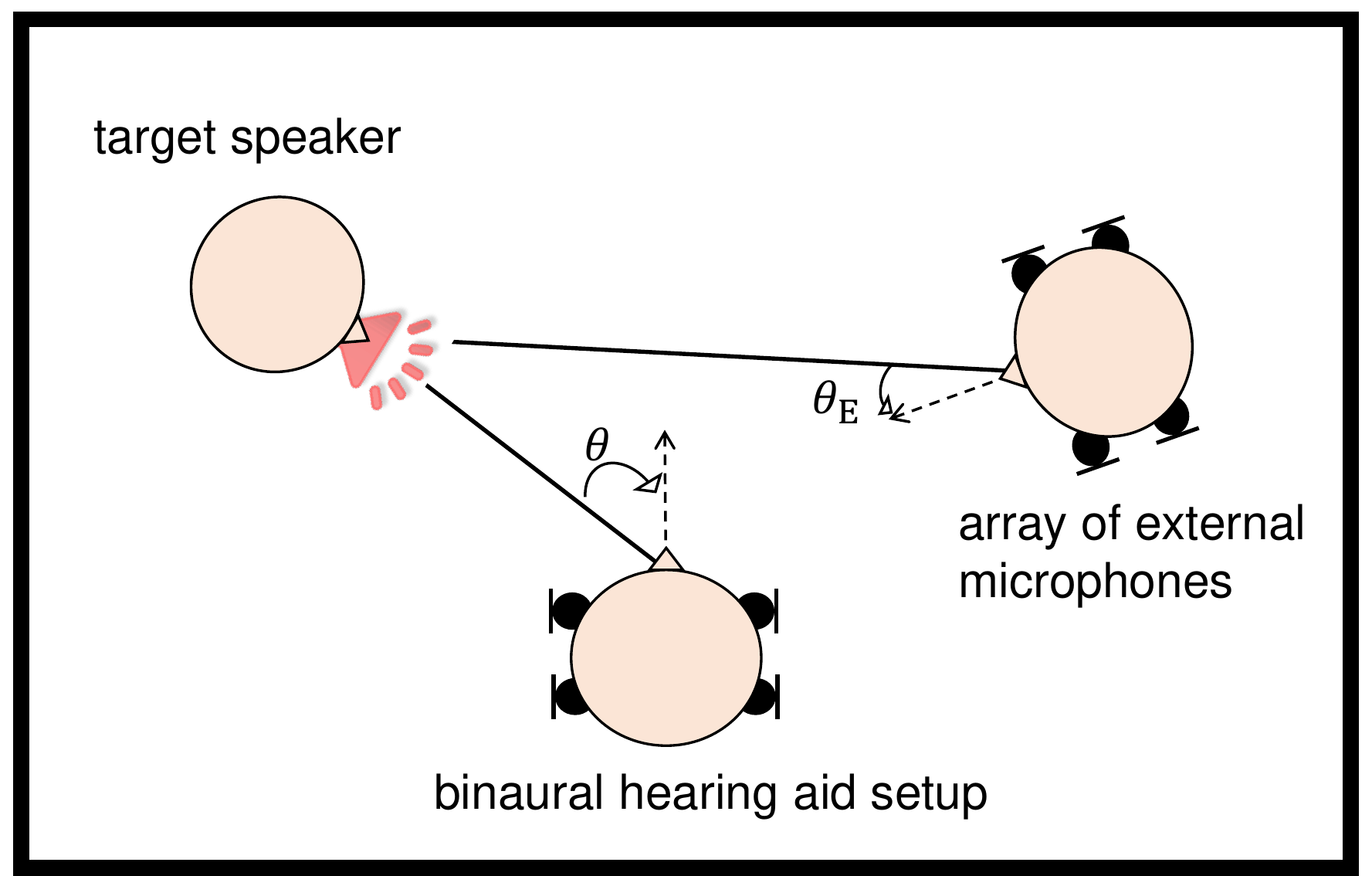}
	\vspace{-3mm}
	\caption{Considered acoustic scenario with one target speaker, a binaural hearing setup and an external microphone array, which in this work corresponds to the same binaural hearing aid setup.\label{fig:setup}}
	\vskip-3.4mm
\end{figure}

Whereas the methods in \cite{Farmani2018,Fejgin2021} exploit a single external microphone, in this paper we assume the availability of a calibrated array of external microphones. With calibrated array we mean that (similarly as for the binaural hearing aid) a database of anechoic prototype RTF vectors for several directions is available for this array. The prototype RTF vectors can be obtained, e.g., through measurement or when the geometry of the external microphone array is known (and assuming free-field sound propagation). It should be noted that the relative position of the external microphone array with respect to the binaural hearing aid setup is obviously unknown

We propose a method where the calibrated array of external microphones assists in estimating the DOA of the target speaker. Similarly as in \cite{Fejgin2021}, the RTF vector between all available microphones and a reference microphone on one of the hearing aids is first estimated using the state-of-the-art covariance whitening method \cite{Markovich2009}. Using the entire RTF vector and the prototype databases of both arrays, i.e. the \mbox{binaural} hearing aid as well as the external microphone array, we now define a two-dimensional spatial spectrum. The DOA of the target speaker is then estimated by determining the location of the main peak of this two-dimensional spatial spectrum. To further investigate the potential of the proposed method, we also consider a special case, by transforming the two-dimensional spatial spectrum into a one-dimensional spatial spectrum using a coordinate system transformation which requires prior information. The performance of the proposed \mbox{binaural} DOA estimation method is evaluated using recordings for a \mbox{single} static speaker in a reverberant acoustic scenario with diffuse-like noise. Experimental results show the benefit of exploiting a calibrated external microphone array compared to exploiting an uncalibrated external microphone array or using only the hearing aid microphones for DOA estimation.

\section{Signal model and notation}
\label{sec:signalModel}
\vskip-1.5ex We consider an acoustic sensor network composed of two microphone arrays with a total of $M = \numLoc+\numEmics$ microphones (see Fig. \ref{fig:setup}): a binaural hearing aid setup consisting of $\numLoc$ head-mounted microphones (i.e., $\numLoc/2$ microphones on each hearing aid) and an array consisting of $\numEmics$ external microphones, which is assumed to be spatially separated from the binaural hearing aid setup. We consider a noisy and reverberant acoustic scenario with a single speaker that is located at DOA $\theta$ relative to the local coordinate system of the binaural hearing setup and angle $\theta_{\mathrm{E}}$ relative to the local coordinate system of the external microphone array (in the azimuthal plane). 

In the short-time Fourier transform (STFT) domain, the $m$-th microphone signal can be written as
\begin{equation}
	Y_{m}\left(k,l\right) = X_{m}\left(k,l\right) + N_{m}\left(k,l\right)\,, \quad m \in \left\{1,\dots,M\right\}\,,
	\label{eq:signalModel_micComponent}
\end{equation}
where $X_{m}\left(k,l\right)$ and $N_{m}\left(k,l\right)$ denote the speech and noise component in the $m$-th microphone signal, respectively, and $k\in\left\{1,\dotsc,K\right\}$ and $l\in\left\{1,\dotsc,L\right\}$ denote the frequency bin index and the frame index, respectively. Since all frequency bins are assumed to be independent and are hence processed independently, the index $k$ will be omitted in the remainder of the paper where possible. Stacking all microphone signals into the vector $\vectorY = \left[Y_{1}\left(l\right),\dotsc,Y_{\numLoc}\left(l\right),\dotsc,Y_{M}\left(l\right)\right]^{T} \in \mathbb{C}^{M}$, where $\left(\cdot\right)^{T}$ denotes transposition, the noisy microphone signals can be written as $\vectorY = \vectorX + \vectorN$, where the speech vector $\vectorX$ and the noise vector $\vectorN$ are defined similarly as $\vectorY$. 

Assuming that the speech vector can be split into a direct-path component $\vectorXdp$ and a re\-ver\-berant component $\vectorXrev$ and \mbox{assuming} that the multiplicative transfer function approximation \cite{Avargel2007} holds for the direct-path component, the speech vector $\vectorX$ can be written as
\begin{equation}
	\vectorX = \vectorXdp + \vectorXrev = 
	\begin{bmatrix}
		\mathbf{a}_{\mathrm{H}}\left(\theta\right)\\
		\mathbf{a}_{\mathrm{E}}\left(\theta_{\mathrm{E}}\right)\\
	\end{bmatrix}S\left(l\right) + \vectorXrev\,,\label{eq:modelSpeechComponentATF}
\end{equation}
where the $\numLoc$-dimensional vector $\mathbf{a}_{\mathrm{H}}\left(\theta\right)$ and the $\numEmics$-dimen\-sional vector $\mathbf{a}_{\mathrm{E}}\left(\theta_{\mathrm{E}}\right)$ denote the direct-path acoustic transfer function vectors between the speaker $S$ and the hearing aid microphones and the external microphones, respectively. By introducing the direct-path RTF vectors $\mathbf{g}_{\mathrm{H}}\left(\theta\right) = \mathbf{a}_{\mathrm{H}}\left(\theta\right)/A_{1}\left(\theta\right)$ and $\mathbf{g}_{\mathrm{E}}\left(\theta_{\mathrm{E}}\right) = \mathbf{a}_{\mathrm{E}}\left(\theta_{\mathrm{E}}\right)/A_{\mathrm{E},1}\left(\theta_{\mathrm{E}}\right)$, where the first microphone of each array is chosen as reference microphone without loss of generality, we can rewrite the direct-path speech component in \eqref{eq:modelSpeechComponentATF} as
\begin{equation}
	\vectorXdp  = 
	\begin{bmatrix}
	A_{1}\left(\theta\right)\mathbf{g}_{\mathrm{H}}\left(\theta\right)\\
	A_{\mathrm{E},1}\left(\theta_{\mathrm{E}}\right)\mathbf{g}_{\mathrm{E}}\left(\theta_{\mathrm{E}}\right)\\
	\end{bmatrix}S\left(l\right)=\mathbf{g}\refSpeechDP\,,\label{eq:modelDPSpeechComponentRTF}
\end{equation}
where the $M$-dimensional vector $\mathbf{g}$ denotes the direct-path RTF vector between all microphones and the reference microphone of the hearing aid and $\refSpeechDP$ denotes the direct-path speech component in the re\-ference microphone of the hearing aid. Condensing the noise and reverberation components into the undesired component $\vectorU = \vectorN + \vectorXrev$, the vector of noisy microphone signals can be written as $\vectorY = \vectorXdp + \vectorU$. 

The RTF vectors $\RTFvecHA$ and $\RTFvecE$ can be both extracted from the RTF vector $\RTFvec$ in \eqref{eq:modelDPSpeechComponentRTF} as
\begin{equation}
	\boxed{\RTFvecHA = \selectionMatrixHA\RTFvec\,, \quad \RTFvecE = \frac{\selectionMatrixE\RTFvec}{\selectionVecE^{T}\selectionMatrixE\RTFvec}}
	\label{eq:extractRTFvec_theo}
\end{equation}
\begin{equation}
	\selectionMatrixHA = \left[\mathbf{I}_{\numLoc\times\numLoc},\mathbf{0}_{\numLoc\times\numEmics}\right]\,,\quad \selectionMatrixE = \left[\mathbf{0}_{\numEmics\times\numLoc},\mathbf{I}_{\numEmics\times\numEmics}\right]\,,
\end{equation}
where $\mathbf{I}_{N\times N}$ denotes an $N\times N$-dimensional identity matrix, $\mathbf{0}_{N\times N^{\prime}}$ denotes an $N\times N^{\prime}$ matrix of zeros, and $\selectionVecE=\left[1,0,\dotsc,0\right]^{T}$ denotes the $\numEmics$-dimensional selection vector. Both for the binaural hearing aid setup as well as for the external microphone array, we assume that a database of anechoic prototype RTF vector is available (referred to as calibrated array). The database for the binaural hearing aid setup is denoted as $\gProtoHA,~i=1,...,I$, while the database for the external microphone array is denotes as $\gProtoE,~ j=1,...,J$.

Assuming that the direct-path speech component is un\-correlated with the undesired component, the $M\times M$-dimensional co\-va\-riance matrix of the noisy microphone signals can be written as
\begin{equation}
	\phiY = \myExpectation{\vectorY\vectorYherm} = \RTFvec \RTFvecherm \Phi_{\rm{x}}^{\rm{DP}}\left(l\right) + \phiU\,,
	\label{eq:def_PhiY}
\end{equation}
where $\left(\cdot\right)^{H}$ and $\myExpectation{\cdot}$ denote complex transposition and expectation operators, respectively, $\Phi_{\rm{x}}^{\rm{DP}}\left(l\right) = \myExpectation{|\refSpeechDP|^{2}}$ denotes the power spectral density of the direct-path speech component in the reference microphone, and $\phiU= \myExpectation{\vectorU\vectorUherm}$ denotes the covariance matrix of the undesired component. The $\numLoc\times\numLoc$-dimensional covariance matrices $\phiYHA$ and $\phiUHA$ corres\-ponding to the noisy microphone signals and the undesired components in the hearing aid microphones can be extracted from \eqref{eq:def_PhiY} as
\begin{equation}
	\phiYHA = \selectionMatrixHA\phiY\selectionMatrixHA^{T}\,, \quad \phiUHA = \selectionMatrixHA\phiU\selectionMatrixHA^{T}\,.
\end{equation}

\section{RTF-vector-based DOA estimation}
\label{sec:DOAest}
\vskip-1.5ex To estimate the DOA $\theta$ of the target speaker relative to the \mbox{binaural} hearing setup, in this section we propose an extension of the RTF-vector-based DOA estimation method from \cite{Fejgin2021}. In Section \ref{ssec:DOAest_noCoop} we review the \mbox{existing} method from \cite{Fejgin2021}, where either no external microphone array or an uncalibrated external microphone array is used. In Section \ref{ssec:DOAest_Coop} we propose a method to jointly use both calibrated arrays, i.e. the binaural hearing aid setup and the external microphone array for DOA estimation.

\subsection{Baseline RTF-vector-based DOA estimation}
\label{ssec:DOAest_noCoop}
\vskip-1ex To estimate the $\numLoc$-dimensional RTF vector $\RTFvecHANOTHETA$ corresponding to the binaural hearing aid setup, we use the state-of-the-art co\-variance whitening (CW) method \cite{Markovich2009} in each time-frequency bin. This RTF vector can be estimated from only the hearing aid microphone signals as
\begin{align}
	\hatRTFvecHACW &=f\left(\hatphiYHA,\hatphiUHA\right)\label{eq:estRTF_CW}\,,\\
	f\left(\tilde{\boldsymbol{\Phi}}_{\rm{y}},\tilde{\boldsymbol{\Phi}}_{\rm{u}}\right)&=\frac{\tilde{\boldsymbol{\Phi}}_{\rm{u}}^{1/2}\myPrincipalEigenvec{\tilde{\boldsymbol{\Phi}}_{\rm{u}}^{-1/2}\tilde{\boldsymbol{\Phi}}_{\rm{y}}\tilde{\boldsymbol{\Phi}}_{\rm{u}}^{-H/2}}}{\tilde{\mathbf{e}}_{1}^{T}\tilde{\boldsymbol{\Phi}}_{\rm{u}}^{1/2}\myPrincipalEigenvec{\tilde{\boldsymbol{\Phi}}_{\rm{u}}^{-1/2}\tilde{\boldsymbol{\Phi}}_{\rm{y}}\tilde{\boldsymbol{\Phi}}_{\rm{u}}^{-H/2}}}\,,
\end{align}
where $\myPrincipalEigenvec{\cdot}$ denotes the principal eigenvector of a matrix, $\tilde{\boldsymbol{\Phi}}_{\rm{u}}^{1/2}$ denotes a square-root decomposition (e.g., Cholesky decomposition) of $\tilde{\boldsymbol{\Phi}}_{\rm{u}}$ and $\tilde{\mathbf{e}}_{1}=\left[1,0,\dotsc,0\right]^{T}$ denotes a selection vector of same dimensionality as the matrix $\tilde{\boldsymbol{\Phi}}_{\rm{u}}$. Alternatively, the RTF vector $\RTFvecHANOTHETA$ can be estimated from all microphone signals, i.e. the hearing aid and external microphone signals, as
\begin{align}
	\hatRTFvecCWE &=f\left(\hatphiY,\hatphiU\right)\,,\label{eq:estRTF_CWE}\\
	\hatRTFvecHACWE &= \selectionMatrixHA\,\hatRTFvecCWE\,.\label{eq:estRTF_CWEHA}
\end{align} 
Due to the involved eigenvalue and square-root decomposition, it can be easily shown that in general $\hatRTFvecHACWE\neq\hatRTFvecHACW$. In \cite{Hassani2015} it was experimentally shown that the RTF vector $\RTFvecHANOTHETA$ can be more accurately estimated using all available microphone signals than using only the hearing aid microphone signals.

Based on the estimated RTF vector $\hatRTFvecHAkl$ corresponding to only the hearing aid microphone signals and the database of anechoic prototype RTF vectors $\gProtoHA$, a one-dimensional spatial spectrum is constructed as
\begin{equation}
	P\left(l,\theta_{i}\right) = -\sum_{k=2}^{K-1}d\left(\hatRTFvecHAkl,~\gProtoHA\right)\,.\label{eq:spatialSpectrum_FejginEUSIPCO2021}
\end{equation}
The DOA of the speaker is estimated as the location of the main peak of this spatial spectrum. It should be noted that the spatial spectrum in \eqref{eq:spatialSpectrum_FejginEUSIPCO2021} is obtained via frequency-averaging of narrowband spatial spectra in order to make the DOA estimation more robust against estimation errors in the RTF vector at individual frequencies. Each narrowband spatial spectrum is obtained by assessing the similarity between the estimated RTF vector and an anechoic prototype RTF vector. \mbox{Inspired} by \cite{Marquardt2018}, in this paper we consider a similarity measure based on the $\ell_{2}$-norm, i.e.
\begin{equation}
	d\left(\mathbf{a},\mathbf{b}\right) = \|\exp\left(i\angle\mathbf{a}\right) - \exp\left(i\angle\mathbf{b}\right)\|_{2}\,,\label{eq:def_PHATL2norm}
\end{equation}
where the operators $\angle\cdot$ and $\exp\left(\cdot\right)$ are applied element-wise to extract the phase and apply the exponential.

\subsection{Assisted RTF-vector-based DOA estimation}
\label{ssec:DOAest_Coop}
\vskip-1ex When considering an uncalibrated external microphone array as in Section \ref{ssec:DOAest_noCoop}, it should be realized that only the database $\gProtoHA$ and thus only the estimated RTF vector $\hatRTFvecHAkl$ correspon\-ding to the hearing aid microphones can be considered for the construction of a one-dimensional spatial spectrum as in \eqref{eq:spatialSpectrum_FejginEUSIPCO2021}. Since in this paper the external microphone array is assumed to be calibrated, i.e. a database $\gProtoE$ of anechoic prototype RTF vectors is available, the entire estimated RTF vector $\hatRTFvec$ and both prototype databases can - and should - be utilized. We first propose to construct a two-dimensional spatial spectrum that exploits both RTF vector databases $\gProtoHA$ and  $\gProtoE$ jointly with the entire RTF vector $\hatRTFvecCWEkl$ in order to exploit spatial correlations between microphone signals of both arrays for improving the DOA estimation. We then consider a special case, requiring prior information about relative position and orientation of both arrays.

\begin{figure}[t]% 
	\centering
	\input{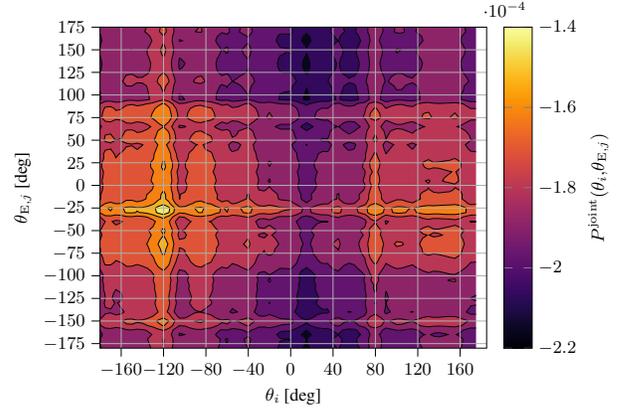}%
	\vspace{-3mm}
	\caption{Example of a two-dimensional spatial spectrum $P^{\rm{joint}}\left(\theta_{i},\theta_{\mathrm{E},j}\right)$. Corresponding acoustic scene: $\theta = -120^{\circ}$ and $\theta_{\mathrm{E}} = -25^{\circ}$, $T_{\rm 60}\approx 1250\mathrm{ms}$, $\mathrm{SNR} = 30\mathrm{dB}$.}
	\vskip-3.4mm
	\label{fig:2SspatialSpectrum}
\end{figure}

We define two $M$-dimensional concatenated RTF vectors, similarly to $\mathbf{g}$ in \eqref{eq:modelDPSpeechComponentRTF}. The vector
\begin{equation}
	\gProtoTwoD = 
	\begin{bmatrix}
		\gProtoHA\\
		\gProtoE
	\end{bmatrix}\,.\label{eq:gProto2D}
\end{equation}
concatenates the anechoic prototype RTF vectors from both databases and is obtained for each pair of DOAs $\theta_{i}$ and $\theta_{\mathrm{E},j}$. Similarly, the vector
\begin{equation}
	\tildeRTFkl = 
	\begin{bmatrix}
		\hatRTFvecHACWEkl\\
		\hatRTFvecECWEkl
	\end{bmatrix}
	\label{eq:tildeRTF_stack}
\end{equation}
concatenates the estimated RTF vector corresponding to the \mbox{hearing} aid microphones $\hatRTFvecHACWEkl$ in \eqref{eq:estRTF_CWEHA} and the estimated RTF vector corresponding to the external microphones, which is equal to
\begin{equation}
	\hatRTFvecECWE =\frac{\selectionMatrixE\,\hatRTFvecCWE}{\selectionVecE^{T}\selectionMatrixE\,\hatRTFvecCWE}
	\label{eq:gEhat}
\end{equation}
similarly to \eqref{eq:extractRTFvec_theo}. A two-dimensional spatial spectrum is now constructed by assessing the similarity between the concatenated vectors in \eqref{eq:tildeRTF_stack} and \eqref{eq:gEhat}, i.e.
\begin{equation}
	\boxed{P^{\rm{joint}}\left(l,\theta_{i},\theta_{\mathrm{E},j}\right) = -\sum_{k=2}^{K-1}d\left(\tildeRTFkl,~\gProtoTwoD\right)}\label{eq:spatialSpectrum_2D}
\end{equation} 
with the similarity measure defined in \eqref{eq:def_PHATL2norm}. The DOA of the speaker is estimated as the location of the main peak of this two-dimensional spatial spectrum. Fig. \ref{fig:2SspatialSpectrum} depicts an exemplary spectrum for a static speaker in a reverberant environment with spatially diffuse babble noise. It should be realized that the main peak consists of the estimated DOA $\hat{\theta}\left(l\right)$ as well as the estimated angle $\hat{\theta}_{\rm{E}}\left(l\right)$ of the speaker as \grqq{}seen\grqq{} from both local coordinate systems (compare Fig. \ref{fig:setup}). In this paper, however, we will evaluate the estimation accuracy of $\hat{\theta}\left(l\right)$ only, as we are mainly interested in how the array of external microphones assists the binaural hearing aid setup in estimating the DOA $\theta$.

To further investigate the potential of the proposed method, we also consider a special case that transforms the two-dimensional spatial spectrum in \eqref{eq:spatialSpectrum_2D} into a one-dimensional spatial spectrum. Contrary to $P^{\rm{joint}}\left(l,\theta_{i},\theta_{\mathrm{E},j}\right)$ that considers each pair of DOAs $\theta_{i}$ and $\theta_{\mathrm{E},j}$ in this special case only a subset of pairs is considered. In particular, only those pairs of DOAs $\theta_{i}$ and $\theta_{\mathrm{E},i}$ are considered where the respective prototype RTF vectors $\gProtoHA$ and $\bar{\mathbf{g}}_{\rm{E}}\left(k,\theta_{\mathrm{E},i}\right)$ are spatially matched, i.e. steer towards the same candidate position. To obtain the respective DOA pairs prior information about the relative position and orientation of both arrays is required. Based on this information, a coordinate system transformation between the two local coordinate systems can be applied, i.e. $\theta_{\mathrm{E},i} = g\left(\theta_{i},\theta_{\mathrm{E},j}\right)$. Enforcing this matching condition makes the resulting spatial spectrum one-dimensional, i.e.
\begin{equation}
	P^{\rm{match}}\left(l,\theta_{i}\right) = -\sum_{k=2}^{K-1}d\left(\tildeRTFkl,\gProto\right)\,.\label{eq:spatialSpectrum_match}
\end{equation}
Similarly to \eqref{eq:spatialSpectrum_2D}, the DOA of the speaker is estimated as the location of the main peak.

\section{Experimental results}
\label{sec:experiments}
\vskip-1.5ex In this section we compare the DOA estimation accuracy when the exploited array of external microphone is calibrated or not. The experimental setup is described in Section \ref{ssec:detailsSetup} and the considered algorithms and implementation details are described in Section \ref{ssec:algos}. The results are presented and
discussed Section \ref{ssec:results}.

\subsection{Experimental setup and implementation details}
\label{ssec:detailsSetup}
\vskip-1ex For the experiments we used signals that were recorded in a la\-boratory at the University of Oldenburg with dimensions about $7\times6\times2.7 \mathrm{m}^{3}$, where the reverberation time can be adjusted by means of absorber pa\-nels, which are mounted to the walls and the ceiling. The reverberation time is set to approxi\-mately $T_{\rm 60}\approx 1250\mathrm{ms}$. A dummy head with a binaural hearing aid setup ($\numLoc$ = 4) is placed approximately in the center of the laboratory. For this hearing aid setup a database of anechoic prototype RTF vectors is obtained from measured anechoic binaural room impulse responses (BRIRs) \cite{Kayser2009} with an angular resolution of $5^{\circ}$ ($I = 72$). As an array of external microphones we consider a second dummy head with the same binaural hearing setup ($\numEmics$ = 4) that is located about $1\mathrm{m}$ and about $80^{\circ}$ to the right side from the first dummy head (see Fig. \ref{fig:setup}). As the same binaural hearing aid setup is con\-sidered for the external microphone array, $\gProtoE = \gProtoHAE$ and $J=I$. The target speaker was a male English speaker played back via a loudspeaker. Relative to the first dummy head 9 different speaker DOAs in the range $\left[-160^{\circ},-120^{\circ},\dotsc,160^{\circ}\right]$ at about $2\mathrm{m}$ distance are considered. The speech signal is constantly active and is approximately $4\mathrm{s}$ long. Diffuse-like noise is generated with four loudspeakers facing the corners of the laboratory, playing back different multi-talker recordings. The speech and noise components are recorded separately and are mixed at $0\mathrm{dB}$ signal-to-noise ratio (SNR) averaged over all microphones of the first dummy head. All microphone signals are assumed to be synchronized such that latency aspects can ne neglected.

\subsection{Algorithms and implementation details}
\label{ssec:algos}
\vskip-1ex We compare the following algorithms to assess whether a calibrated external microphone array can assist the binaural hearing setup in estimating the speaker DOA $\theta$. The notation \grqq{}X/Y\grqq{} means that the microphone array X is used to estimate an RTF vector and that the microphone array Y is used to construct a spatial spectrum, where either the binaural hearing aid setup (denoted as H) alone or jointly with the external microphone array (denoted as H+E) are considered as options.
\begin{itemize}
	\item H/H: considering only the hearing aid microphones of the first dummy head, both to estimate the RTF vector $\hatRTFvecHAkl$ in \eqref{eq:estRTF_CW} as well as to construct the spatial spectrum $P\left(l,\theta_{i}\right)$ in \eqref{eq:spatialSpectrum_FejginEUSIPCO2021}.
	\item H+E/H: considering all microphones to estimate the RTF vector $\hatRTFvecHAkl$ in \eqref{eq:estRTF_CWEHA} but considering only the hearing aid microphones of the first dummy head to construct the spatial spectrum $P\left(l,\theta_{i}\right)$ in \eqref{eq:spatialSpectrum_FejginEUSIPCO2021}.
	\item H+E/H+E (2D): considering all microphones, both to estimate the concatenated RTF vector $\tildeRTFkl$ in \eqref{eq:tildeRTF_stack} as well as to construct the two-dimensional spatial spectrum $P^{\rm{joint}}\left(l,\theta_{i},\theta_{\mathrm{E},j}\right)$ in \eqref{eq:spatialSpectrum_2D}.
	\item H+E/H+E (match): as a special case of H+E/H+E (2D) also all microphones are considered, both to estimate the concatenated RTF vector $\tildeRTFkl$ in \eqref{eq:tildeRTF_stack} as well as to construct the one-dimensional spatial spectrum $P^{\rm{match}}\left(l,\theta_{i}\right)$ in \eqref{eq:spatialSpectrum_match} exploiting prior knowledge about the microphone configuration. Due to the enforced matching condition the spatial spectrum $P^{\rm{match}}\left(l,\theta_{i}\right)$ consists of only 20 candidate speaker positions. 
\end{itemize}

The microphone signals (sampling frequency $f_{\rm s} = 16\mathrm{kHz}$) are processed in the STFT domain using $32\mathrm{ms}$ square-root Hann windows with $50\%$ overlap. For each time-frequency (TF) bin the covariance matrix of the noisy microphone signals $\phiYkl$ and the covariance matrix of the undesired components $\phiUkl$ are estimated recursively during speech-and-noise TF bins and noise-only TF bins as
\begin{align}
	\hatphiYkl &= \alpha_{\rm y}\hatphiYklprev + \left(1-\alpha_{\rm y}\right)\mathbf{y}\left(k,l\right)\mathbf{y}^{H}\left(k,l\right)\,,\label{eq:SOS_RyUpdate}\\
	\hatphiUkl &= \alpha_{\rm u}\hatphiUklprev + \left(1-\alpha_{\rm u}\right)\mathbf{y}\left(k,l\right)\mathbf{y}^{H}\left(k,l\right)\,,\label{eq:SOS_RuUpdate}
\end{align}
where the smoothing factors $\alpha_{\rm y}$ and $\alpha_{\rm u}$ correspond to time constants of $250\mathrm{ms}$ and $500\mathrm{ms}$, respectively. The speech-and-noise TF bins are discriminated from noise-only TF bins based on estimated speech presence probabilities \cite{Gerkmann2012} in the microphones of the first dummy head, which are averaged and thresholded.

To assess the DOA estimation performance, we average the localization accuracy over the considered $9$ speaker DOAs, where the localization accuracy is defined as the percentage of frames that are correctly localized within a range of $\pm5^{\circ}$.

\subsection{Results}
\label{ssec:results}
\begin{figure}[t]% 
	\centering
\definecolor{mycolor1}{rgb}{0.00000,0.44700,0.74100}%
\begin{tikzpicture}[scale=0.195]

\begin{axis}[%
width=15.5in,
height=8.175in,
at={(2.6in,1.103in)},
scale only axis,
bar shift auto,
xmin=-0.2,
xmax=5.2,
xtick={{1},{2},{3},{4}},
xticklabels={{H/H},{H+E/H},{H+E/H+E (2D)},{H+E/H+E (match)}},
xticklabel style={font=\color{black}, font = \fontsize{40}{1}\selectfont,align=center,text width=7.5cm,yshift=-0.5cm},
xlabel style={font=\color{black}, font = \fontsize{40}{1}\selectfont},
ymin=50,
ymax=100,
ytick={{0},{10},{20},{30},{40},{50},{60},{70},{80},{90},{100}},
yticklabels={{0},{10},{20},{30},{40},{50},{60},{70},{80},{90},{100}},
yticklabel style={font=\color{black}, font = \fontsize{40}{1}\selectfont},
ylabel style={font=\color{black}, font = \fontsize{40}{1}\selectfont},
ylabel={Accuracy [\%]},
axis background/.style={fill=white},
xmajorgrids,
ymajorgrids,
legend style={at={(0.01,0.4)}, anchor=north west, legend cell align=left, align=left, draw=none, font = \fontsize{40}{1}\selectfont,row sep=3pt}
]
\addplot[ybar, bar width=0.8, fill=mycolor1, draw=black, area legend, fill opacity=0.75] table[row sep=crcr] {%
1	77.0498084291188\\
2	79.8467432950192\\
3	82.911877394636\\
4	97.7011494252874\\
};
\addplot [draw=red,fill=red, opacity=0.2]
coordinates {(0.5,0) (0.5,100) (2.5,100) (2.5,0)};
\addplot [draw=ForestGreen,fill=ForestGreen, opacity=0.2]
coordinates {(2.5,0) (2.5,100) (4.5,100) (4.5,0)};
%\node[text width=15cm,font=\fontsize{45}{45}\selectfont] at (1.65,65) {\textbf{no or uncalibrated \\ external microphone array}};
%\node[text width=15cm,font=\fontsize{45}{45}\selectfont] at (3.68,66.8) {\textbf{calibrated external microphone array}};
\end{axis}
\end{tikzpicture}%
%
	\vspace{-3mm}
	\caption{Average localization accuracy for the investigated algorithms that either exploit calibrated external microphones (green background) or not (red background).}
	\vskip-3.4mm
	\label{fig:res}
\end{figure}
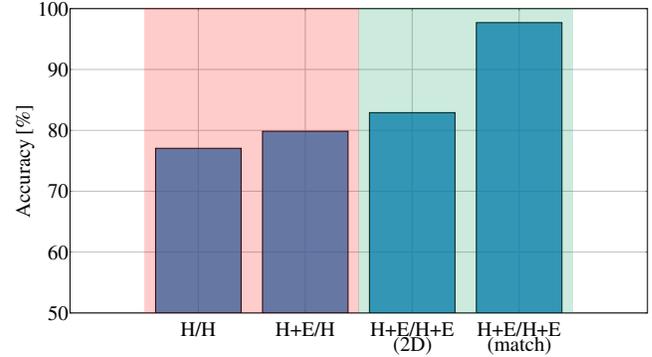
\vskip-1ex Fig. \ref{fig:res} depicts the average localization accuracy for the four investigated algorithms. First, it can be observed that for the algorithm H/H using only the hearing aid microphones of the first dummy head and for the algorithm H+E/H, where the external microphones are exploited only to estimate the RTF vector, there is only a minor performance difference (average localization accuracy for both algorithms about $80\%$). This result is in line with the high-reverberation results reported in \cite{Fejgin2021}. Second, considering the \grqq{}H+E/H+E\grqq{} algorithms, which both exploit the external microphones to estimate the RTF vector as well as to construct a spatial spectrum, the results clearly show that both algorithms yield a larger average localization accuracy than the algorithms that do not consider external microphones to construct a spatial spectrum. In case of the H+E/H+E (2D) algorithm, where a two-dimensional spatial spectrum is constructed, this improved average localization accuracy is due to exploitation of the additional database of prototype RTF vectors $\gProtoE$. Third, in case of the H+E/H+E (match) algorithm, where prior knowledge about the microphone configuration is additionally exploited, the average localization accuracy can be significantly improved. Based on these results, the potential of assisted DOA estimation is clearly demonstrated.

\section{Conclusions}
\label{sec:print}

In this paper we explored how calibrated external microphone arrays can be exploited to assist in estimating the DOA of a target speaker. We extended a recently proposed binaural RTF-vector-based DOA estimation method that considered only uncalibrated external microphone arrays to exploit calibrated external microphone arrays. We proposed to exploit the availability of two databases of anechoic prototype RTF vectors, i.e. the binaural hearing aid as well as the external microphone, in order to construct a two-dimensional spatial spectrum from which the DOA of the speaker can be estimated. We compared RTF-vector-based DOA estimation algorithm that either exploit a calibrated array of external microphones or not. Experimental results clearly demonstrate a benefit of exploiting a calibrated external microphone array compared to not exploiting a calibrated external microphone array or using only the hearing aid microphones for DOA estimation. Thus, the benefit of assisted DOA estimation is clearly demonstrated.

\vfill\pagebreak

\bibliographystyle{IEEEbib}
\bibliography{refs}

\end{document}